\documentclass{ws-procs975x65}

\def\Th{\Theta}
\def\sig{\sigma}

\def\3nab{\tilde{\nabla}}

\def\hsp5{\hspace{5mm}}
\newcommand{\sfrac}[2]{{\textstyle{#1\over#2}}}
\def\case#1/#2{\textstyle\frac{#1}{#2}}

\def\be {\begin{equation}}
\def\ee {\end{equation}}
\def\bea {\begin{eqnarray}}
\def\eea {\end{eqnarray}}

\def\bi {\bibitem}
\def\case#1/#2{\textstyle\frac{#1}{#2} }
\def\rf#1{(\ref{#1})}
\def\cqg{{\em Class. Quantum Grav.\/} }

\def\prd{{\em Phys. Rev.\/} {\bf D}}
\def\prl{{\em Phys. Rev. Lett.\/} }
\def\apj{{\em Astrophys. J.\/} }

\begin{document}

\title{AN ANALYSIS OF THE SHEAR DYNAMICS IN BIANCHI I \\ COSMOLOGIES WITH $R^n$-GRAVITY}

\author{J. A. LEACH$^{1\,*}$, P. K. S. DUNSBY$^{1\,2}$ and S. CARLONI$^{1}$}

\address{1. Department of Mathematics and Applied Mathematics,\\ University of Cape Town, Rondebosch,
7701, South Africa\\
2. South African Astronomical Observatory,\\ Observatory, Cape Town,
South Africa \\ $^*$E-mail: jannie.leach@uct.ac.za\\}

\begin{abstract}
We consider the case of $R^n$-gravity and perform a detailed
analysis of the dynamics in Bianchi I cosmologies which exhibit {\it
local rotational symmetry}  (LRS). We find exact solutions and study their
behaviour and stability in terms of the values of the parameter $n$.
In particular, we found a set of cosmic histories in which the
universe is initially isotropic, then develops shear anisotropies
which approaches a constant value.
\end{abstract}


\bodymatter

\section{Introduction}

The dynamical systems approach \cite{Dynamical} has been used with great success over the last 30 years,  to  a gain
(qualitative) description of the global dynamics of cosmological models. This method provides a useful tool for finding exact
solutions which correspond to fixed points of the system. 

Carloni {\it et al} \cite{Carloni05} have recently used this method
to study the dynamics of $R^n$-theories in
Friedmann-Lema\^{i}tre-Robertson-Walker (FLRW) universes. Clifton
and Barrow \cite{Clifton05} used the dynamical systems approach to
determine the extent to which exact solutions can be considered as
attractors of spatially flat universes at late times. They compared
the predictions of these results with a range of observations and
argued that the parameter $n$ in FLRW may only deviate
from GR by a very small amount ($n-1\sim 10^{-19}$).

The main aim of this paper \cite{Leach06} is to see how the shear behaves in LRS
Bianchi I cosmologies in $R^n$- gravity and whether these models
isotropises at early and late times. To achieve this goal we use
the theory of dynamical systems \cite{Dynamical} to analyse the system of equations
governing the evolution of this model with and without matter.

\section{Dynamics in LRS Bianchi I cosmologies}

The cosmological equations that we require for our analysis are:
\begin{eqnarray}
&&\dot{\Th} + \sfrac{1}{3}\,\Th^{2} + 2\sigma^2 -\frac{1}{2n}R -
(n-1)\frac{\dot{R}}{R}\Th+\frac{\mu}{nR^{n-1}} =0,
\label{Ray:R^n_B1} \\
&&
\sfrac{1}{3}\,\Th^{2}-\sigma^2+(n-1)\frac{\dot{R}}{R}\Th-\frac{(n-1)}{2n}R-\frac{\mu}{nR^{n-1}}=0,
\label{3R:Rn_B1} \\
&&\dot{\sig}= -\left(\Th+ (n-1)\frac{\dot{R}}{R}\right)\sigma,
\label{sigdot:Rn_LRS_B1} \\
&&\dot{\mu}=-(1+w)\mu\Th, \label{cons:perfect}
\end{eqnarray}
where $\Theta =\sfrac{3\dot{a}}{a}$ is the volume expansion and
$\sigma$ is the shear
($\sigma^2=\sfrac{1}{2}\sigma_{ab}\sigma^{ab}$).

In order to convert the above equations  into a system of autonomous
first order differential equations, we define the following set of
expansion normalised variables;
\begin{equation}\label{DS:var}
\Sigma =\frac{3\sigma^2}{\Th^2}\; , \quad x =
\frac{3\dot{R}}{R\Th}(n-1)\; ,\quad  y = \frac{3R}{2n\Th^2}(n-1)\; ,
\quad  z = \frac{3\mu}{nR^{n-1}\Th^2}\; ,
\end{equation} whose
equations are
\begin{eqnarray}\label{DS:eqn_mat}
&& \Sigma'=2\left(-2+2\Sigma-\frac{y}{n-1}-2x+z\right)\Sigma,
\nonumber
\\
&& x' = y(2+x)-\frac{y}{n-1}(2+nx)-2x-2x^2+xz+(1-3w)z+2x\Sigma,  \\
&& y'=\frac{y}{n-1}\left[(3-2n)x-2y+2(n-1)z+4(n-1)\Sigma+2(n-1)\right], \nonumber \\
&& z' = z\left[ 2z-(1+3w)-3x-\frac{2y}{n-1}+4\Sigma\right] ,
\nonumber\\
&& 1-\Sigma+x-y-z=0, \nonumber
\end{eqnarray}
where primes denote derivatives with respect to a new time variable
$\tau=\ln a$.

The solutions associated to the fixed points can be obtained from:
\begin{eqnarray}
&& \dot{\Th}=\frac{1}{3 \alpha}\Th^2, \quad \quad \quad \quad
\alpha=\left(2+\Sigma_i-\sfrac{n}{n-1}y_i\right)^{-1},
\label{Theta_eqn} \\
&& \frac{\dot{\sig}}{\sigma}= - \frac{\beta}{3}\Th, \quad \quad
\quad \quad \beta =2+\Sigma_i+y_i+z_i. \label{sigma_eqn}
\end{eqnarray}
Under the condition that $n \neq 1$ and the terms inside the
brackets of \rf {Theta_eqn}, do not add up to zero, these equations
may be integrated to give the following solutions
\begin{equation}\label{DS:sol_gen_mat}
a=a_0\left(t-t_0\right)^\alpha, \quad \quad \quad \quad
\sigma=\sigma_0 a^{-\beta}.
\end{equation}

In terms of our expansion normalised variables
\rf{DS:var}, the energy density  is given by
\begin{equation}\label{DS:mu}
\mu \propto zy^{n-1}\Th^{2n}.
\end{equation}
From this relation it can be seen that when $z=0$ and $y \neq 0$ the
energy density is zero. However when $y=0$ and $z \neq 0$ the
behaviour of $\mu$ does depend on the value of $n$. In this case the
energy density is zero when $n>1$ but is divergent when $n<1$. When
both $y$ and $z$ are equal to zero and $n<1$, one can only determine
the behaviour of $\mu$ by direct substitution into the cosmological
equations.
\subsection{Vacuum analysis}

The vacuum case is characterised  by  $z=0$. In this case we can obtain the fixed points of   \rf{DS:eqn_mat} by setting  $\Sigma'=0$ and $y'=0$.  We find one isotropic
fixed point ($\mathcal{A}):\ \left(0,\sfrac{4n-5}{2n-1}\right)$ and a
line of fixed points ($\mathcal{L}_1):\ \left(\Sigma_*,0\right)$, with non-vanishing shear. The isotropic fixed
point is an attractor (stable node) for values of the parameter $n$
in the ranges $n<1/2$, $1/2<n<1$ and $n>5/4$. In the range $1<n<5/4$
this point is a repeller (unstable node) and therefore may be seen
as a past attractor.
However, we also have attractors for $(\sigma/H)_*<<1$ on
$\mathcal{L}_1$. Therefore inflation may not be needed since the shear
anisotropy approaches a constant value which may be chosen as the
expansion normalised shear observed today\cite{Bunn96,Kogut97,Jaffe05} ($(\sigma/H)_*< 10^{-9}$), provided that other observational
constraints such as nucleosynthesis are satisfied.

The phase space is divided into two regions by the line $y=1-\Sigma$, which represents all points for which the shear dissipate at the same rate as in GR. The region $y<1-\Sigma$ represents a {\it fast shear
dissipation} (FSD) regime where shear dissipates faster than in
GR, and the region $y>1-\Sigma$, is a {\it slow shear
dissipation} (SSD) regime where the shear dissipates
slower than in GR (see Figure 1). 

\def\figsubcap#1{\par\noindent\centering\footnotesize(#1)}
\begin{figure}[tbp]%
\begin{center}
  \parbox{2.1in}{\epsfig{figure=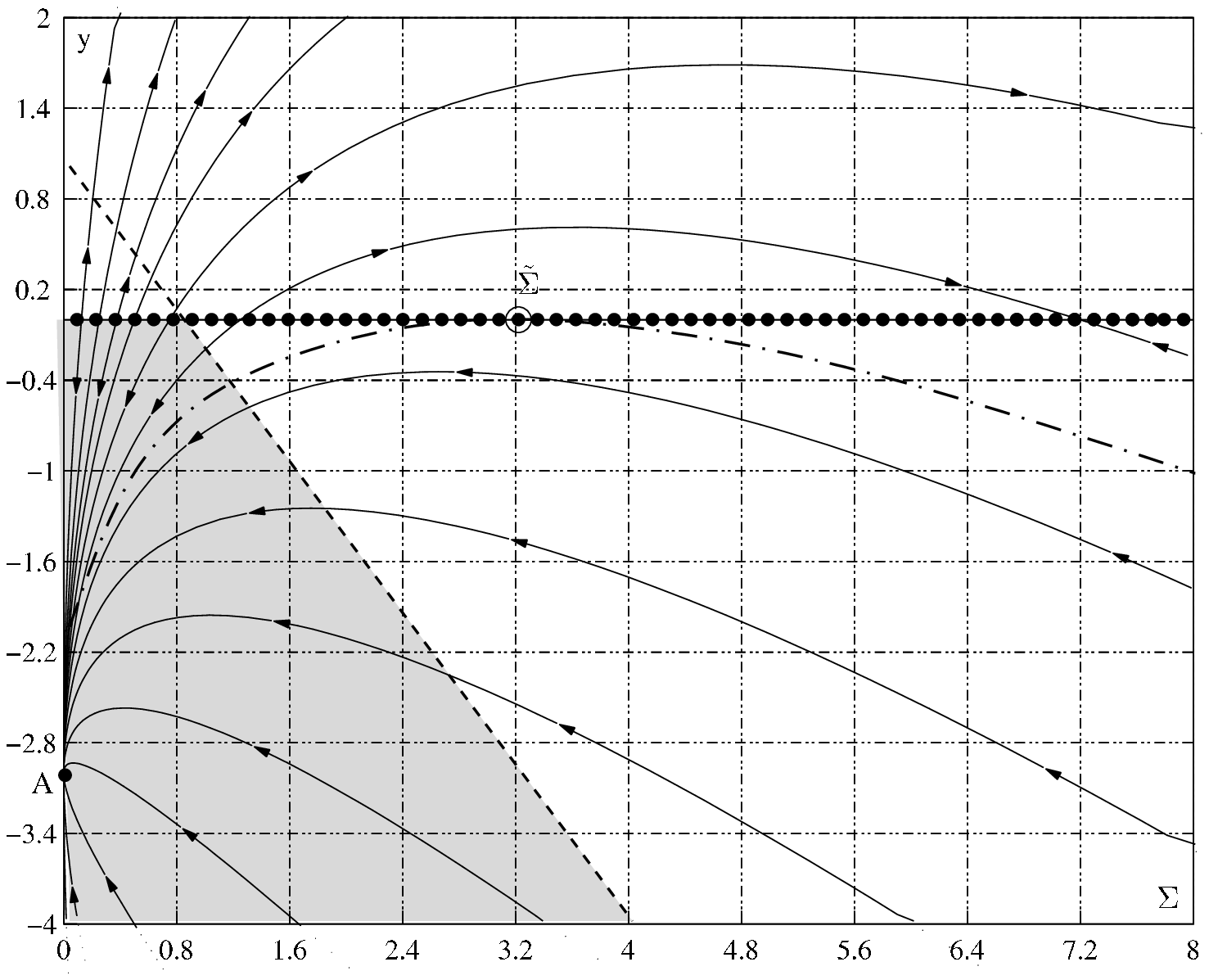,width=2 in}\figsubcap{a}}
  \hspace*{4pt}
  \parbox{2.1in}{\epsfig{figure=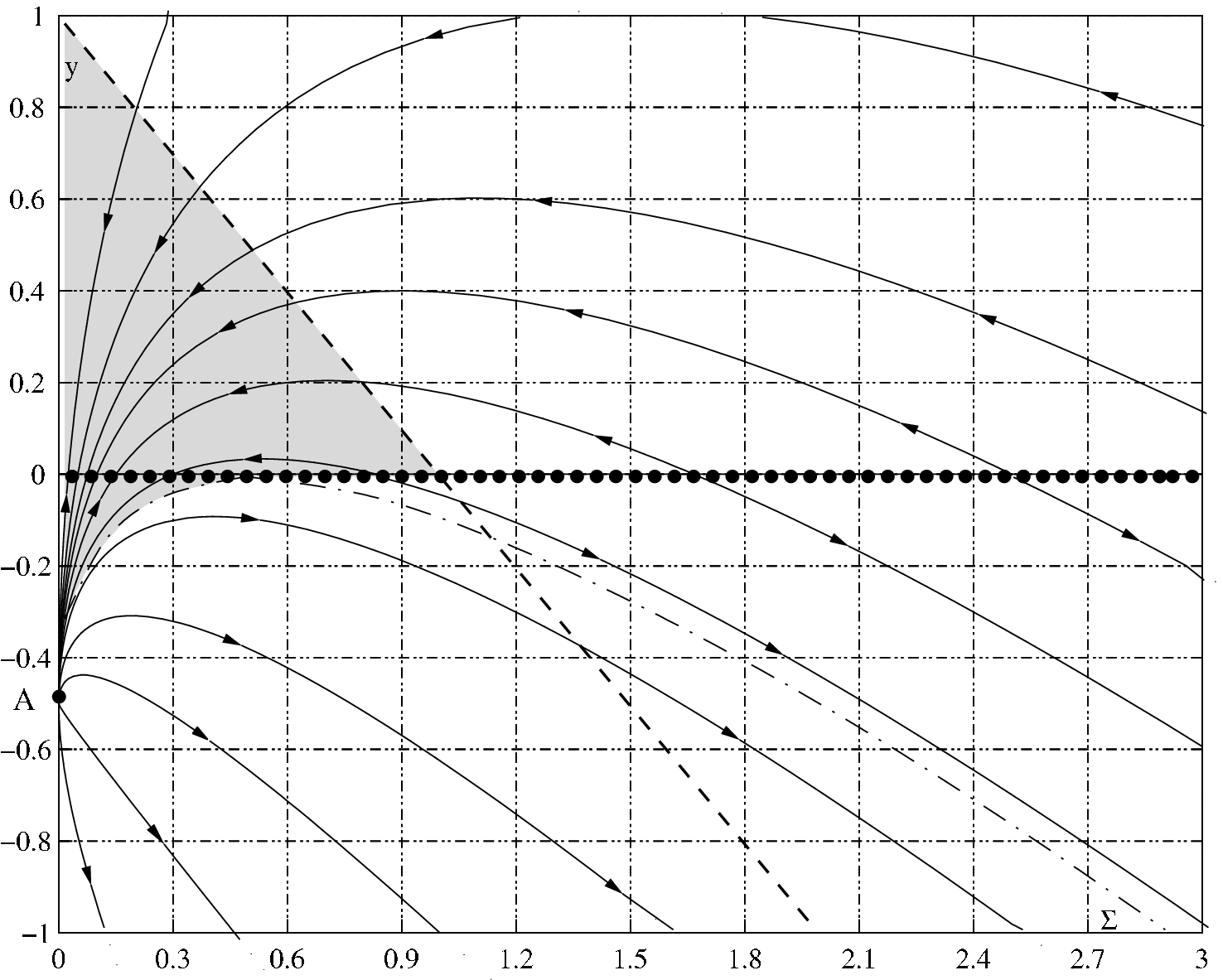,width=2 in}\figsubcap{b}}
  \caption{Phase space of the
vacuum model. (a) $1/2<n<1$. (b) $1<n<5/4$. The shaded regions
represents the regions of initial conditions for which the shear
will always evolve
faster than in GR.}%
  \label{fig1}
\end{center}
\end{figure}

\subsection{Matter analysis}
Setting $\Sigma'=0$, $y'=0$ and $z'=0$ we obtain three isotropic
fixed points and a line
of fixed points with non-vanishing shear.

We observe the same kind of behaviour as in the vacuum case;  the
phase space is however 3-dimensional, but is similarly divided into
two regions, by the plane $1=\Sigma+y+z$. The space above the plane
is a SSD region and below a FSD region.

\section{Conclusions}

In conclusion we have shown that $R^n$- gravity modifies the
dynamics of the shear in LRS Bianchi I cosmologies by altering the
rate at which the shear dissipates. There are cases in which the
shear always dissipates slower or faster than in GR, and there are
ones which make the transition from first evolving faster and then
slower (and {\it vice versa}) than in GR.

\end{document}